\documentstyle[12pt]{article}

\expandafter\ifx\csname mathrm\endcsname\relax\def\mathrm#1{{\rm #1}}\fi

\makeatletter
\if@twoside \oddsidemargin -17pt \evensidemargin 00pt
\else \oddsidemargin 00pt \evensidemargin 00pt
\fi
\makeatother

\def\beq{\begin{equation}}
\def\eeq{\end{equation}}
\def\beqar{\begin{eqnarray}}
\def\eeqar{\end{eqnarray}}
\def\barr#1{\begin{array}{#1}}
\def\earr{\end{array}}
\def\bfi{\begin{figure}}
\def\efi{\end{figure}}
\def\btab{\begin{table}}
\def\etab{\end{table}}
\def\bce{\begin{center}}
\def\ece{\end{center}}
\def\nn{\nonumber}

\def\text{\textstyle}
\def\arraystretch{1.2}

\def\al{\alpha}

\def\ga{\gamma}
\def\de{\delta}
\def\eps{\varepsilon}

\def\si{\sigma}

\def\Ga{\Gamma}
\def\De{\Delta}

\def\gWpm{g_{\PWpm}^2(0)}
\def\gWO{g_{\PWO}^2(\MZ^2)}

\def\refeq#1{\mbox{(\ref{#1})}}
\def\reffi#1{\mbox{Fig.~\ref{#1}}}
\def\reffis#1{\mbox{Figs.~\ref{#1}}}
\def\refta#1{\mbox{Tab.~\ref{#1}}}

\def\refse#1{\mbox{Sect.~\ref{#1}}}

\def\refapp#1{\mbox{App.~\ref{#1}}}
\def\citere#1{\mbox{Ref.~\cite{#1}}}
\def\citeres#1{\mbox{Refs.~\cite{#1}}}

\def\solid{\raise.9mm\hbox{\protect\rule{1.1cm}{.2mm}}}
\def\dash{\raise.9mm\hbox{\protect\rule{2mm}{.2mm}}\hspace*{1mm}}

\newcommand{\GeV}{\unskip\,\mathrm{GeV}}
\newcommand{\MeV}{\unskip\,\mathrm{MeV}}
\newcommand{\TeV}{\unskip\,\mathrm{TeV}}

\newcommand{\ord}{{\cal O}}

\newcommand{\Oa}{\mathswitch{{\cal{O}}(\alpha)}}

\def\mathswitchr#1{\relax\ifmmode{\mathrm{#1}}\else$\mathrm{#1}$\fi}

\newcommand{\PW}{\mathswitchr W}
\newcommand{\PZ}{\mathswitchr Z}

\newcommand{\PH}{\mathswitchr H}
\newcommand{\Pe}{\mathswitchr e}

\newcommand{\Pl}{\mathswitchr l}

\newcommand{\Pt}{\mathswitchr t}

\newcommand{\PWpm}{\mathswitchr {W^\pm}}
\newcommand{\PWO}{\mathswitchr {W^0}}

\newcommand{\Gl}{\Ga_{\mathrm l}}

\def\mathswitch#1{\relax\ifmmode#1\else$#1$\fi}

\newcommand{\MW}{\mathswitch {M_\PW}}
\newcommand{\MWpm}{\mathswitch {M_\PWpm}}
\newcommand{\MWO}{\mathswitch {M_\PWO}}

\newcommand{\MZ}{\mathswitch {M_\PZ}}
\newcommand{\MH}{\mathswitch {M_\PH}}

\newcommand{\Mt}{\mathswitch {m_\Pt}}

\renewcommand{\ss}{\scriptscriptstyle}
\newcommand{\sw}{\mathswitch {s_{\ss\PW}}}

\newcommand{\swbar}{\mathswitch {\bar s_{\ss\PW}}}

\newcommand{\GF}{\mathswitch {G_\mu}}

\newcommand{\logx}{f_1}
\newcommand{\logy}{f_2}
\newcommand{\alpz}{\alpha(\MZ^2)}
\newcommand{\alps}{\alpha_{\mathrm s}}

\newcommand{\bos}{{\mathrm{bos}}}
\newcommand{\fer}{{\mathrm{ferm}}}

\newcommand{\SC}{{\mathrm{SC}}}
\newcommand{\IB}{{\mathrm{IB}}}

\newcommand{\LEP}{{\mathrm{LEP}}}

\def\Li{\mathop{\mathrm{Li}_2}\nolimits}
\def\Re{\mathop{\mathrm{Re}}\nolimits}

\def\draftdate{\relax}
\def\mda{\relax}
\def\mua{\relax}
\def\mla{\relax}
\def\draft{
\def\thtystars{******************************}
\def\sixtystars{\thtystars\thtystars}
\typeout{}
\typeout{\sixtystars**}
\typeout{* Draft mode!
         For final version remove \protect\draft\space in source file *}
\typeout{\sixtystars**}
\typeout{}
\def\draftdate{\today}
\def\mua{\marginpar[\boldmath\hfil$\uparrow$]%
                   {\boldmath$\uparrow$\hfil}%
                    \typeout{marginpar: $\uparrow$}\ignorespaces}
\def\mda{\marginpar[\boldmath\hfil$\downarrow$]%
                   {\boldmath$\downarrow$\hfil}%
                    \typeout{marginpar: $\downarrow$}\ignorespaces}
\def\mla{\marginpar[\boldmath\hfil$\rightarrow$]%
                   {\boldmath$\leftarrow $\hfil}%
                    \typeout{marginpar: $\leftrightarrow$}\ignorespaces}
\def\Mua{\marginpar[\boldmath\hfil$\Uparrow$]%
                   {\boldmath$\Uparrow$\hfil}%
                    \typeout{marginpar: $\Uparrow$}\ignorespaces}
\def\Mda{\marginpar[\boldmath\hfil$\Downarrow$]%
                   {\boldmath$\Downarrow$\hfil}%
                    \typeout{marginpar: $\Downarrow$}\ignorespaces}
\def\Mla{\marginpar[\boldmath\hfil$\Rightarrow$]%
                   {\boldmath$\Leftarrow $\hfil}%
                    \typeout{marginpar: $\Leftrightarrow$}\ignorespaces}
\overfullrule 5pt
\oddsidemargin -15mm
\marginparwidth 29mm
}

\makeatletter

\def\eqnarray{\stepcounter{equation}\let\@currentlabel=\theequation
\global\@eqnswtrue
\global\@eqcnt\z@\tabskip\@centering\let\\=\@eqncr
$$\halign to \displaywidth\bgroup\hskip\@centering
  $\displaystyle\tabskip\z@{##}$\@eqnsel&\global\@eqcnt\@ne
  \hskip 2\arraycolsep \hfil${##}$\hfil
  &\global\@eqcnt\tw@ \hskip 2\arraycolsep $\displaystyle\tabskip\z@{##}$\hfil
   \tabskip\@centering&\llap{##}\tabskip\z@\cr}
\def\appendix{\par
 \setcounter{section}{0} \setcounter{subsection}{0}
 \def\thesection{\Alph{section}}}

\makeatother

\begin{document}
\thispagestyle{empty}
\def\thefootnote{\fnsymbol{footnote}}
\setcounter{footnote}{1}
\null
\hfill BI-TP 95/31 \\
\null
\hfill hep-ph/9510386
\vskip .8cm
\begin{center}
{\Large \bf On the Observation of Bosonic Loop Corrections in
Electroweak Precision Experiments}
\footnote{Partially supported by the EC-network contract CHRX-CT94-0579.}
\vskip 3em
{\large S.\ Dittmaier, D.\ Schildknecht and G.\ Weiglein%
\footnote{Supported by the Bundesministerium f\"ur Bildung und Forschung,
Bonn, Germany.}% \\
}
\vskip .5em
{\it Fakult\"at f\"ur Physik, Universit\"at Bielefeld, Germany}
\vskip 2em
\end{center} \par
\vskip 1.2cm
\vfil

{\bf Abstract} \par
We investigate the structure of the experimentally significant electroweak
bosonic loop corrections to the leptonic weak mixing angle, the leptonic
Z-boson decay width, and the $\PW$-boson mass.
It is shown that the bosonic corrections that have a sizable
effect at the
present level of experimental accuracy are directly
related to the use of the Fermi constant $\GF$ as input parameter
for analyzing the LEP observables. 
Indeed, if the (theoretical value of the) leptonic width of the W~boson
is used as input parameter instead of the low-energy parameter $\GF$
determined from muon decay, 
fermion-loop corrections are sufficient
for compatibility between theory and
experiment.
\par
\vskip 1cm
\noindent October 1995 \par
\null
\setcounter{page}{0}
\clearpage
\def\thefootnote{\arabic{footnote}}
\setcounter{footnote}{0}

\section{Introduction}
\label{intro}
The data taken at the Z-boson resonance at LEP1 and the determination of
the W-boson mass provide the most stringent test of the electroweak
Standard Model (SM) at present. 
As previously emphasized~\cite{GounSchi}, genuine precision
tests of the electroweak theory require an experimental accuracy that
allows to distinguish between the pure fermion-loop and the full
one-loop predictions of the theory. This accuracy was first reached in
1994. Indeed, by systematically discriminating between fermion-loop
(vacuum-polarization) corrections to the 
photon, $\PZ$- and $\PWpm$-boson
propagators and the full one-loop results, it was found~\cite{zph1,zph2}
that contributions beyond fermion loops are required for consistency
with the experimental results on the leptonic $\PZ$-peak observables and the 
$\PW$-boson mass. While the pure fermion-loop predictions were shown to
be incompatible with the data on the leptonic $\PZ$-boson decay width
$\Ga_\Pl$, the effective weak mixing angle $\bar\sw^2$, and $\MWpm$,
the complete one-loop prediction of the
SM provides a consistent description of the 
experimental results.
Consequently, the data have become sensitive to 
the non-Abelian gauge structure of the standard electroweak theory
entering the bosonic radiative corrections.

The investigations
performed in \citeres{zph1,zph2} differ from related work~\cite{okun}
that 
is concerned with the evidence for radiative corrections beyond the
$\alpz$-Born approximation.
While the $\alpz$-Born approximation contains fermion-loop corrections to
the photon propagator only, in \citeres{zph1,zph2} the full fermion-loop
predictions are compared 
with the experimental data and the complete SM result, thus 
exploring the nature of the electroweak loop corrections and
revealing that in
fact {\it bosonic\/} electroweak corrections are required for consistency with
the data. The experimental evidence for bosonic loop corrections was
also explored for the single observable $\bar\sw^2$ in \citere{ga94}
and for $\MWpm$ in \citere{hi95}.

The necessity of taking into account standard bosonic corrections 
for consistency with the data thus being established, 
it is desirable to explore the detailed structure of those bosonic
corrections to which the experiments are actually sensitive. This is the
topic of the present article.
In order to investigate the structure of the radiative corrections it is
convenient to employ a suitable
set of effective parameters. In \citere{zph1} three effective
parameters, $\De x$, $\De y$, and $\eps$, have been used to accommodate
the radiative corrections to the 
observables $\Ga_\Pl$, $\bar\sw^2$, and $\MWpm$.
These parameters quantify 
possible sources of SU(2) violation in the framework
of an effective Lagrangian~\cite{zph0} for electroweak interactions.
They are directly related to observables and are thus manifestly
gauge-independent quantities.
This analysis was extended in \citere{zph2} by including also the
hadronic decay modes of the $\PZ$~boson.
It has been shown that 
the experimentally resolved bosonic corrections are entirely contained
in the single parameter $\De y$ (corresponding to $\eps_2$ in the
notation of \citere{alta}), which in turn is extremely insensitive to
variations in the Higgs-boson mass.

The parameter $\De y$ relates the effective
charged-current coupling, $g_\PWpm(0)$, determined from muon decay
via the Fermi constant $\GF$, to its neutral-current counterpart,
$g_\PWO (\MZ^2)$, appearing in Z-boson decay.
Thus, $\De y$ quantifies isospin breaking between the neutral-current
and charged-current interactions as well as the transition from the
low-energy process muon decay to the energy scale of the LEP1
observables, i.e.\ the Z-boson mass. 

The isospin-breaking effect 
in the $(W^{\pm},W^0)$ triplet will be studied in \refse{sec:dely}
of the present paper by relating Z-boson decay
to $\PWpm$-boson decay. 
It will be shown that the bosonic contribution to the
isospin-breaking effect is small in the standard electroweak theory when
compared with present (and even future) experimental accuracy. 
As a consequence,
the bosonic corrections significant for the current precision
experiments are related to the transition from the 
charged-current coupling $g_\PWpm(0)$ defined via muon decay to the
charged-current coupling at the scale of the W-boson mass,
$g_\PWpm(\MWpm^2)$, being deduced from the W-boson decay width into
leptons.
It will be demonstrated that it is indeed sufficient
to combine the
fermion-loop predictions with the bosonic contribution
furnishing the transition from $g_\PWpm(0)$ to $g_\PWpm(\MWpm^2)$
in order to achieve agreement between theory and experiment. The quality
of this approximation is illustrated not only at the level of the
effective parameters but also directly for the leptonic LEP1 observables
and the W-boson mass. 

In \refse{sec:gawscheme} it will be shown that the occurrence of the
experimentally relevant bosonic corrections could completely be avoided
by introducing the W-boson decay width as 
input parameter for analyzing the precision data
instead of the low-energy quantity $\GF$.
Using the SM theoretical value of the leptonic W-boson width as
input, it will explicitly be demonstrated that %in such an analysis 
omission of the standard bosonic corrections to the LEP1 observables and
the W-boson mass does not lead to a significant
deviation between theory and experiment. 
Final conclusions are drawn in \refse{conc}.
The appendix provides some auxiliary formulae.

\section{\boldmath{Scale-change and isospin-breaking contributions to
the parameter $\De y$}}
\label{sec:dely}
\subsection{Definitions}
\label{sec:def}

We work with the effective Lagrangian ${\cal L} = {\cal L}_{\mathrm C} +
{\cal L}_{\mathrm N}$ for LEP1 observables introduced in
\citeres{zph1,zph2,zph0}. In the leptonic sector SU(2)-breaking effects
are quantified by the parameters $x$, $y$, $\eps$. More precisely, $x$
is defined as the mass ratio 
in the $(W^{\pm},W^0)$
triplet, 
\beq
\MWpm^2 = x\MWO^2 = (1+\De x) \MWO^2,
\label{mwpm}
\eeq
$y$ as the ratio 
of the effective $\PWpm$ and $\PWO$ couplings to charged leptons, 
\beq
\gWpm = y \gWO = (1+\De y) \gWO,
\label{y}
\eeq
and $\eps$ quantifies SU(2) 
violation in $\ga W^0$ mixing,
\beq
{\cal L}_{\mathrm mix} = -\frac{1}{2} \frac{e(\MZ^2)}{g_\PWO(\MZ^2)}
(1-\eps) A_{\mu \nu} W^{0,\mu \nu}.
\label{lm}
\eeq

In this section we focus on the parameter $y$, which
is the only one incorporating significant bosonic corrections.
The charged-current 
Lagrangian
${\cal L}_{\mathrm C}$ has the form
\beq
{\cal L}_{\mathrm C}=-{1\over 2} W^{+\mu \nu} W^-_{\mu \nu} -
\frac{g_\PWpm}{\sqrt 2}\left( j^+_\mu W^{+\mu} + h.c.\right)
+ \MWpm^2 W^+_\mu W^{-\mu}.
\label{lc}
\eeq
In \citeres{zph1,zph2, zph0} the charged-current coupling $g_\PWpm$
in \refeq{lc} was
defined with respect to muon decay, i.e.\ at a low-energy scale, as 
\beq
\label{eq:gw0}
g^2_\PWpm(0) \equiv 4\sqrt{2}\GF\MWpm^2 .
\eeq
The choice of the low-energy quantity $g_\PWpm(0)$ is of course 
dictated by the fact that $\GF$ is the most accurately known electroweak
parameter apart from the LEP observables to be analyzed.
According to \refeq{y}, the parameter $\De y$ describes
both the transition from the charged-current to the neutral-current
sector and the change from 
the low-energy process muon decay to the energy scale of the LEP
observables.

The effective coupling of the $W^0$~field to charged leptons, 
$g_\PWO(\MZ^2)$, is derived from Z-boson decay, i.e.\ from the 
neutral-current process where all associated particles are physical 
(on-shell). In order to separately study the effect of the
isospin-breaking transition to the charged-current sector, one therefore
has to consider the corresponding (charged-current) process of the
isospin partner of the $W^0$~field, namely leptonic decay of the
$\PWpm$~boson.
Accordingly, we introduce the charged-current coupling at the W-boson
mass shell, $g_\PWpm(\MWpm^2)$, 
which is derived from the leptonic width $\Ga^{\PW}_{\mathrm l}$ of the
W~boson, 
\beq
g^2_\PWpm(\MWpm^2) \equiv \frac{48 \pi}{\MWpm} \Ga^{\PW}_{\mathrm l}
\left(1 + c_0^2 \frac{3 \al}{4 \pi} \right)^{-1} ,
\label{eq:Wlepwidth}
\eeq
where $c_0^2$ is defined~\cite{zph0} according to
\beq
c_0^2 s_0^2 \equiv c_0^2 (1 - c_0^2) = \frac{\pi \al(\MZ^2)}{\sqrt{2}
\GF \MZ^2} .
\label{eq:c0def}
\eeq
In analogy to~\refeq{y} we relate $g_\PWpm(\MWpm^2)$ to 
$g_\PWO (\MZ^2)$ by a parameter $\De y^{\IB}$,
\beq
\label{eq:delyib}
g^2_\PWpm(\MWpm^2) = y^{\IB} g^2_\PWO(\MZ^2)
= (1+\De y^{\IB})g^2_\PWO (\MZ^2) ,
\eeq
where the index ``IB'' refers to weak ``isospin-breaking''.
In \refeq{eq:Wlepwidth} we have introduced a factor 
$(1 + c_0^2 3 \al/(4 \pi) )$ by convention. It is related to the
convention chosen in the treatment of the photonic corrections to the
leptonic $\PZ$-boson decay width $\Gl$~\cite{zph1,zph2,zph0}. The
photonic contributions to $\Gl$ are pure QED corrections giving rise to a
factor $(1 + 3 \al/(4 \pi))$ that is split off and not
included in $\De x$, $\De y$, and $\eps$. 
For the decay of the $\PW$~boson, however,
it is not possible to uniquely separate the QED contribution
from the other one-loop
corrections. As isospin breaking is associated with electromagnetic
interactions, a meaningful definition of $\De y^{\IB}$ %necessarily
requires to treat the photonic corrections on the same footing in both
the neutral and charged vector boson decay. One possibility to achieve
this would be to %not at all split off a QED factor in both the
include all photonic contributions into the bosonic corrections both for
$\PZ$-boson and $\PW$-boson decay. Equivalently, as
far as the magnitude of $\De y^{\IB}$ is concerned, one may keep the
convention for the correction factor 
$(1+3\alpha/(4\pi))$
in the leptonic
$\PZ$-boson decay width $\Gl$ and split off
the corresponding factor also in the decay width of the W~boson, 
$\Ga^\PW_\Pl$.
This is the procedure adopted in~\refeq{eq:Wlepwidth}. The appearance of
$c_0^2$ in the correction factor 
in the W-boson decay width is due to the
rotation in isospin space relating the
physical field 
$Z$ to the field $W^0$ entering the SU(2)
isotriplet. Numerically the correction term introduced in
\refeq{eq:Wlepwidth} amounts to $c_0^2 3\al/(4 \pi)=1.3\times 10^{-3}$. 
Even though the 
introduction of this correction term in \refeq{eq:Wlepwidth} is
well justified, it is worth noting that a different treatment of the
photonic corrections, such as omission of the correction factor in
\refeq{eq:Wlepwidth}, 
would only lead to minor changes that 
would not influence our final conclusions at all.

In the language of our effective Lagrangian~\refeq{lc}, the
transition from the charged-current coupling at the scale of the
muon mass, $g^2_\PWpm(0)$, to the charged-current coupling obtained
from the decay of the W~boson into leptons, $g^2_\PWpm(\MWpm^2)$, 
is denoted as a scale change effect 
and is expressed by a parameter $\De y^{\SC}$,
\beq
g^2_\PWpm(0) = y^{\SC}g^2_\PWpm(\MWpm^2)
= (1+\De y^{\SC})g^2_\PWpm(\MWpm^2) ,
\label{eq:delysc}
\eeq
where the index ``SC'' means ``scale change''. 
Inserting \refeq{eq:delyib} into \refeq{eq:delysc} and comparing with
\refeq{y}, one finds that
in linear approximation the parameter $\De y$ is
split into two additive contributions,
\beq
\label{eq:dely2}
\De y = \De y^{\SC} + \De y^{\IB} ,
\eeq 
which furnish the transition from $g^2_\PWpm(0)$ to $g^2_\PWpm(\MWpm^2)$
and from $g^2_\PWpm(\MWpm^2)$ to $g^2_\PWO (\MZ^2)$, respectively.
Upon substituting \refeq{eq:gw0} and \refeq{eq:Wlepwidth} in
\refeq{eq:delysc}, one finds %for $\De y^{\SC}$
\beq
\label{eq:GaWySC}
\Delta y^{\SC} = 
\frac{\MWpm^3\GF}{6\sqrt{2}\pi\Ga^{\PW}_{\mathrm l}} - 1
+c_0^2 \frac{3 \al}{4 \pi} ,
\eeq
which allows to determine $\Delta y^{\SC}$ 
(and consequently also $\Delta y^{\IB}$ from $\De y$ according
to~\refeq{eq:dely2})
both experimentally and theoretically.

In our analysis of the observables $\Ga_\Pl$, $\bar\sw^2$, and
$\MWpm$, which are measured at the LEP energy scale,
the influence of the low-energy input parameter
$\GF$ determined from muon decay is localized in the single effective
parameter $\De y^\SC$. In an investigation performed directly at the
level of the observables, on the other hand, the effect of using this
low-energy input parameter for analyzing high-energy observables cannot
be separated from the other corrections.

The phrase ``scale change'' used for $\De y^\SC$ should not be confused
with an ordinary ``running'' of universal (propagator-type)
contributions. The couplings $g_\PWpm(0)$ and $g_\PWpm(\MWpm^2)$, being
defined with reference to muon decay and $\PW$-boson decay,
respectively, are obviously process-dependent quantities. 
While the fermion-loop contributions to $\De y^\SC$ 
are of propagator-type, the bosonic contributions to $\De y^\SC$ contain
self-energy, vertex and box corrections that cannot uniquely be
separated on physical grounds.
As all our effective parameters are directly related to 
specific observables, 
i.e.\ to complete S-matrix elements, they are 
manifestly gauge-independent. 

\subsection{Predictions in the Standard Model}

In order to derive the Standard Model prediction for $\Delta y^{\mathrm
SC}$, we have evaluated the one-loop corrections to the leptonic decay 
$\PW \to \Pl \bar\nu_{\Pl}$, $(\Pl = \Pe, \mu, \tau)$, where the
leptons and light quarks are treated as massless and the
bremsstrahlung corrections integrated over the full phase space 
are included in the width.
For the calculation of the radiative corrections we have applied
standard techniques, which are e.g.\ reviewed in \citere{adhabil}.
We have checked that our result for the \Oa\ corrections to 
$\Ga^{\PW}_{\mathrm l}$ is in agreement with the 
result of \citere{de90},
where the \Oa\ corrected $\Ga^{\PW}_{\mathrm l}$ is given for arbitrary 
fermion masses.

We obtain for the 
fermion-loop contribution
to $\Delta y^{\SC}$
\beqar
\label{eq:delYSCferm}
\Delta y^{\SC}_{\fer} &=& 
\left.\Re\left(\frac{\Sigma_{\mathrm T, ferm}^{W}(p^2)-
\Sigma_{\mathrm T, ferm}^{W}(\MWpm^2)}
{p^2 - \MWpm^2}\right)\right|^{p^2 \to \MWpm^2}_{p^2=0} \nn\\
&=& \frac{\al(\MZ^2)}{16 \pi s_0^2 c_0^6}
\left[c_0^2 (6 t^2 + 3 c_0^2 t - 16 c_0^4) 
+ 6 t (t^2 - c_0^4) \log\left(1 - c_0^2/t\right) \right] ,
\eeqar
where $\Sigma_{\mathrm T, ferm}^{W}(p^2)$ denotes the fermion-loop
contribution to the transverse part of the 
unrenormalized $\PW$-boson
self-energy, and the shorthand $t=\Mt^2/\MZ^2$ is introduced. 
As can be seen from \refeq{eq:delYSCferm},
the fermionic contributions to $\De y^{\SC}$ are entirely given as the
difference of the vacuum polarization at $p^2 = \MWpm^2$ and at $p^2=0$.
It does not
give rise to $\log\left(m_\Pt\right)$ terms in the limit of a heavy 
top-quark mass.
For $t\to\infty$ the fermionic contribution $\Delta y^{\SC}_{\fer}$ 
is entirely given by the constant part arising from light fermion doublets,
\beq
\Delta y^{\SC}_{\fer}({\mathrm dom}) \equiv \Delta y^{\SC}_{\mathrm
ferm} \biggr|_{t \to \infty} = - \frac{3 \al(\MZ^2)}{4 \pi s_0^2} 
\sim - 8.01 \times 10^{-3},
\label{eq:delyscfermasy}
\eeq
which
reflects the fact that 
the scale change between zero-momentum squared and $\MWpm^2$ becomes
irrelevant for the contribution of an infinitely heavy top quark in the
loop, i.e.~the top quark decouples in the scale-change contribution.
The negative sign of $\Delta y^{\SC}_{\fer}$,
according to the definition \refeq{eq:delysc},
shows that in
analogy to the QED case the fermion loops lead to an increase of
$g_\PWpm(\MWpm^2)$ relative to $g_\PWpm(0)$.

The 
fermion-loop contribution to $\Delta y^{\IB}$ can directly be
obtained from \refeq{eq:dely2} of the previous section
and formulae (19), (A.1) of
\citere{zph1}. 
We give the dominant term in the limit
of a heavy top quark,
\beq
\Delta y^{\IB}_{\fer}({\mathrm dom}) 
= \frac{\al(\MZ^2)}{8 \pi s_0^2}
\Bigl[\log(t) + 6 \log(c_0^2) - \frac{1}{2} + \frac{40 s_0^2}{3}
- \frac{160 s_0^4}{9} \Bigr] ,
\label{eq:delyibfermasy}
\eeq
which approximates $\Delta y^{\IB}_{\fer}$ up to terms of 
${\cal O}(1/\Mt^2)$.
Numerically we have $\Delta y^{\IB}_{\fer}({\mathrm dom}) = 1.89
\times 10^{-3}$ for $\Mt = 180 \GeV$.
In distinction to \refeq{eq:delYSCferm}, the effect of
isospin breaking increases
logarithmically with $\Mt$.

In \refta{ta:delY}
we give numerical results for the fermionic and bosonic
contributions to $\De y$, $\De y^{\SC}$, and $\De y^{\IB}$
for different values of $m_\Pt$ and $\MH$. 
Table~\ref{ta:delY} shows that
for reasonable values of the top-quark mass the 
asymptotic expansions
\refeq{eq:delyscfermasy} and \refeq{eq:delyibfermasy} obtained for an
infinitely heavy top quark agree with the 
exact results for $\Delta
y^{\SC}_{\fer}$ and $\Delta y^{\IB}_{\fer}$ within $1
\times 10^{-3}$,
justifying the terminology ``dominant''.

\btab 
$$\begin{array}{|c||c|c|c|}
\hline
\Mt/\GeV & \De y_{\fer}/10^{-3} &
\De y^{\SC}_{\fer}/10^{-3} &
\De y^{\IB}_{\fer}/10^{-3} \\ \hline\hline
120 & -7.57 & -7.42 & -0.15 \\ \hline
180 & -6.27 & -7.79 & \phantom{-}1.52 \\ \hline
240 & -5.44 & -7.90 & \phantom{-}2.46 \\ \hline
\earr$$
$$\begin{array}{|c||c|c|c|}
\hline
\MH/\GeV & \De y_{\bos}/10^{-3} &
\De y^{\SC}_{\bos}/10^{-3} &
\De y^{\IB}_{\bos}/10^{-3} \\ \hline\hline
100 & 13.72 & 12.47 & 1.25 \\ \hline
300 & 13.62 & 12.42 & 1.20 \\ \hline
1000 & 13.61 & 12.41 & 1.20 \\ \hline
\earr$$
\caption{Fermionic and bosonic contributions to $\De y$, $\De y^{\SC}$,
and $\De y^{\IB}$ for different values of $\Mt$ and $\MH$.}
\label{ta:delY}
\etab

We turn to 
the bosonic contributions $\Delta y^{\SC}_{\bos}$ and $\Delta
y^{\IB}_{\bos}$ to $\De y$. They
are insensitive to variations in the Higgs-boson mass $\MH$, as in
particular they do not contain a $\log\left(\MH\right)$ 
term for large $\MH$.
The absence of a $\log\left(\MH\right)$ contribution in 
the sum $\De y=\De y^\SC+\De y^\IB$ was 
analyzed in \citere{di95} where it was shown that the heavy-Higgs limit
of $\De y$ in the SM coincides with the prediction in the Higgs-less
(non-renormalizable) {\it massive vector-boson theory\/} 
(i.e.\ the SU(2)$\times$U(1)
{\it gauged non-linear $\si$-model\/}) which corresponds to the SM in the
unitary gauge without physical Higgs field. 
The lack of a $\log(\MH)$ term in $\De y$ can also
be understood from the (custodial)
SU(2)$_{\mathrm C}$ symmetry of the SM. Even though loop corrections do not
necessarily 
respect this symmetry, SU(2)$_{\mathrm C}$-breaking terms generated by
loops with a heavy Higgs-boson are nevertheless suppressed 
by a factor $1/\MH^2$ relative to naive dimensional analysis. This
result can be read off from the general effective one-loop Lagrangian
\cite{he94sdcgk} which quantifies the difference between the SM with a
heavy Higgs boson and the massive vector-boson theory.
For the SU(2)$_{\mathrm C}$-breaking parameter $\De y$ 
the $1/\MH^2$ suppression
implies the absence of a $\log(\MH)$ term, in 
distinction from $\eps$ which
corresponds to an SU(2)$_{\mathrm C}$-conserving interaction and 
contains a $\log(\MH)$ term,
although both $\De y$ and $\eps$ are related to dim-4 interactions.
The SU(2)$_{\mathrm C}$-violating
parameter $\De x$ behaves like $\log(\MH)$, as the $\MH^2\log(\MH)$
term naively expected for this dim-2 interaction term is absent.

Consequently, in the limit $\MH\to\infty$
the bosonic contribution to $\Delta y^{\SC}$ has the constant value
\beqar
\Delta y^{\SC}_{\bos}({\mathrm dom}) &\equiv &
\Delta y^{\SC}_{\bos}
\biggr|_{\MH^2 \to \infty} 
\nn\\[.3em] 
&=& \frac{\al(\MZ^2)}{16 \pi s_0^2}
\left[\frac{1}{6 c_0^4} (18 - 81 c_0^2 + 157 c_0^4 + 296 c_0^6 
+ 32 c_0^4 s_0^2 \pi^2) 
- 16 (2 + c_0^2) C_7 
\right. \nn\\ && {} 
+ \frac{4}{c_0^6} (1 - 2 c_0^2) (1 + c_0^2)^2 C_6
+ \frac{1}{2 c_0^6} (1 + 13 c_0^2 - 52 c_0^4 - 28 c_0^6) f_2(c_0^2)
\nn\\ && \left. {} 
- \frac{1}{2 c_0^6 s_0^2} (1 + 18 c_0^2 - 103 c_0^4 + 94 c_0^6 + 36 c_0^8
- 40 c_0^{10}) \log(c_0^2) 
\right] + c_0^2 \frac{3 \al}{4 \pi}.
\hspace{2em}
\label{eq:scbosdom}
\eeqar
The definitions of the function $f_2(x)$ and the abbreviations 
$C_6$ and $C_7$ are given in \refapp{aux}.
Numerically \refeq{eq:scbosdom} amounts to
\beq
\Delta y^{\SC}_{\bos}({\mathrm dom}) = 12.41 \times 10^{-3} .
\label{eq:scbosdomnum}
\eeq
Comparison of \refeq{eq:scbosdomnum} with the
values of $\Delta y^{\SC}_{\bos}$ for non-asymptotic values of
$\MH$ in \refta{ta:delY} shows that 
$\Delta y^{\SC}_{\bos}({\mathrm dom})$
is sufficiently accurate for all practical
purposes. For completeness, we nevertheless give the numerically
irrelevant terms of $\ord(1/\MH^2)$ in the appendix.

As can be seen from \refeq{eq:delyscfermasy}, \refeq{eq:scbosdomnum}
and \refta{ta:delY},
the fermionic and bosonic corrections to $\Delta y^{\SC}$ enter with
different signs. 
This leads to strong cancellations in $\De y^\SC$ and $\De y$.

The analytic result for the isospin-breaking contribution
$\Delta y^{\IB}_{\bos}$ 
can be obtained using 
formulae (20), 
(22), (A.2) of \citere{zph1}
and (33)--(38) of \citere{zph2}.
In the limit of an infinitely heavy
Higgs-boson mass it reads explicitly
\beqar
\Delta y^{\IB}_{\bos}({\mathrm dom}) &\equiv &
\Delta y^{\IB}_{\bos} \biggr|_{\MH^2 \to \infty} 
\nn\\[.3em] &=& 
\frac{\al(\MZ^2)}{16 \pi s_0^2} \left[
-\frac{2s_0^2}{3c_0^4}(5+9c_0^2-140c_0^4+278c_0^6+120c_0^8+8c_0^4\pi^2) 
-\frac{16}{c_0^2}(1-2c_0^2)^3C_1
\right. \nn\\ && \left. {}
-8(1+c_0^2)^2C_2
-16c_0^4(2+c_0^2)C_3
-\frac{4}{c_0^6}(1+c_0^2)^2(1-2c_0^2)C_6
+16(2+c_0^2)C_7
\right. \nn\\ && \left. {}
-(1+26c_0^2-20c_0^4-40c_0^6)\logx(c_0^2)
-\frac{1}{6c_0^6}(2+23c_0^2-88c_0^4-36c_0^6)\logy(c_0^2)
\right. \nn\\ && \left. {}
+\frac{1}{6c_0^6}(2+43c_0^2-150c_0^4-60c_0^6+72c_0^8)\log(c_0^2)
\right]
-c_0^2\frac{3\al}{4\pi}
\nn\\[.3em] 
&=& 1.20\times 10^{-3}.
\label{eq:ibbosdom}
\eeqar
The Higgs-mass dependent remainder part of $\Delta y^{\IB}_{\bos}$,
which is explicitly given in \refapp{sec:remain}, 
is again numerically completely negligible 
(see \refta{ta:delY}).

It is worth noting that our investigation of
$\De y^{\SC}$, as a by-product, has lead to a compact expression for the SM
one-loop result 
for the leptonic $\PW$-boson decay width
$\Ga^{\PW}_{\mathrm l}$. 
According to \refeq{eq:GaWySC}, $\Ga^{\PW}_{\mathrm l}$ is expressed in terms
of $\De y^{\SC}$ and $\MWpm$, 
\beq
\label{eq:GaW}
\Ga^\PW_\Pl = 
\frac{\MWpm^3\GF}{6\sqrt{2}\pi (1+\De y^\SC)}
\left(1+c_0^2\frac{3\al}{4\pi}\right),
\eeq
where the SM prediction for 
$\De y^{\SC}$ is given in \refeq{eq:delYSCferm}, \refeq{eq:scbosdom},
\refeq{eq:delyscbosrem}, and the one for $\MWpm$ 
in terms of $\De x$, $\De y$, $\eps$ is given in
\citeres{zph1,zph2}.

\subsection{Comparison with the experimental data}
\label{sec:discu}

\begin{figure}
\begin{center}
\begin{picture}(16,12)
\put(-2.56,-14.2){\includegraphics{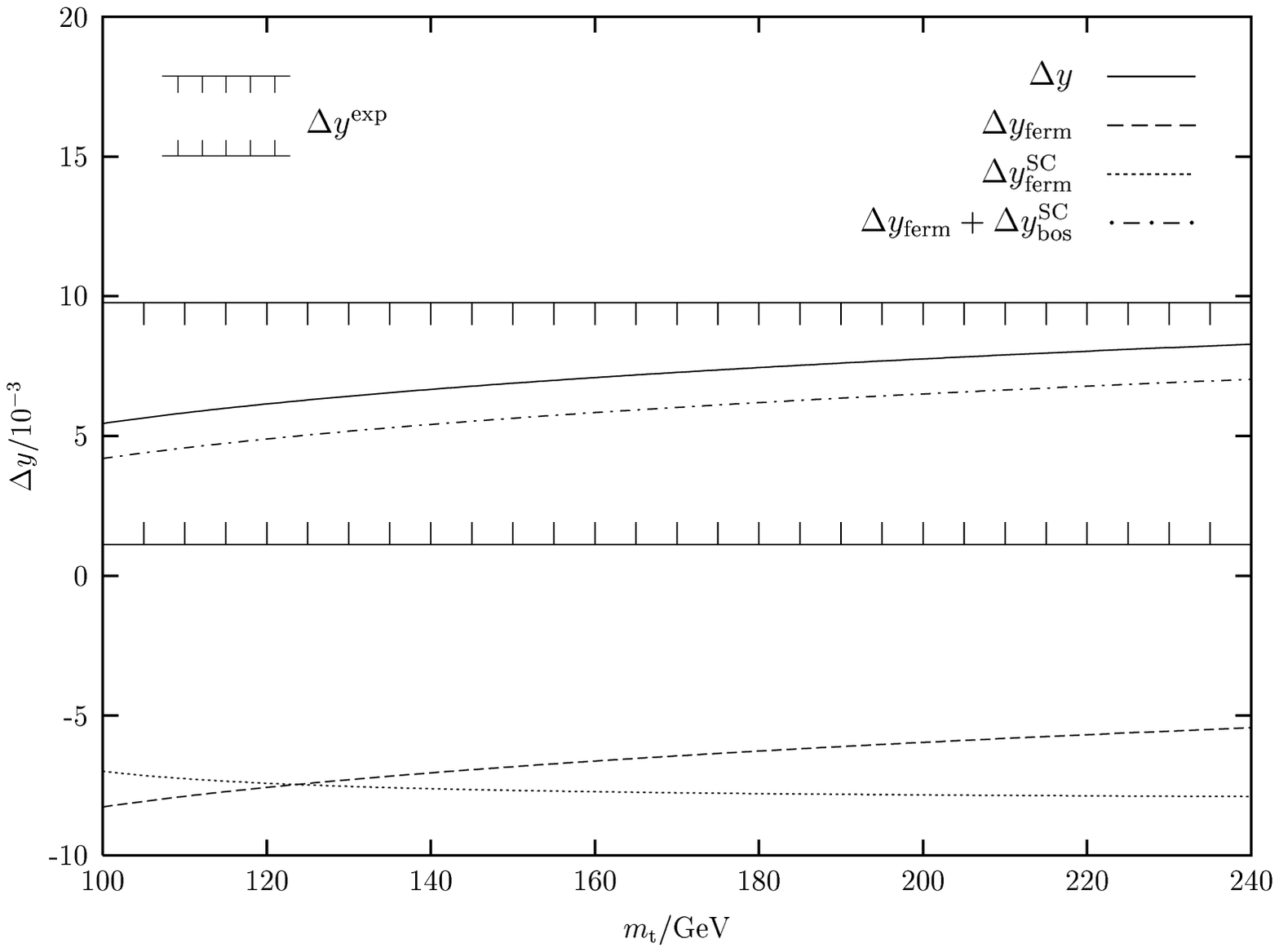}}
\end{picture}
\end{center}
\caption{The one-loop SM predictions for $\De y$, $\De
y_{\fer}$, $\De y^{\SC}_{\fer}$, and $(\De y_{\mathrm
ferm} + \De y^{\SC}_{\bos})$ as a function of $m_\Pt$.
The difference between the curves for $\De y$ and $(\De y_{\fer} +
\De y^{\SC}_{\bos})$ corresponds to the small contribution of $\De
y^{\IB}_{\bos}$.
The experimental value of $\Delta y$,
$\Delta y^{\mathrm exp} = (5.4 \pm 4.3) \times 10^{-3}$, is indicated
by the error band.}
\label{fig:delysc}
\efi

In \reffi{fig:delysc} the theoretical result for $\De y$
and the predictions for various contributions to $\De y$
are compared with the experimental value of $\Delta y$, 
$\Delta y^{\mathrm exp} = (5.4 \pm 4.3) \times 10^{-3}$.
The experimental value of $\Delta y$ is based on the data 
presented for the observables $\bar\sw^2$, $\Gl$, $\MWpm/\MZ$ 
at the 1995 Summer Conferences~\cite{data8/95} 
(see \refeq{scdata} below)
and is obtained
according to the method described in \citere{zph0}. 

Fig.~\ref{fig:delysc} demonstrates that the bosonic and also the
fermionic isospin-breaking contributions, i.e.\ both the difference
between $\De y_{\fer}$ and $\De y^{\SC}_{\fer}$ as well as the
difference between $\De y$ and $(\De y_{\fer} + \De y^{\SC}_{\bos})$,
are considerably smaller than the experimental error of $\De y$.
The pure fermion-loop contribution to $\De y$, as previously
observed~\cite{zph1,zph2}, is inconsistent with the experimental data.
For full agreement with the data it is sufficient to add the bosonic
scale-change contribution, $\De y^{\SC}_{\bos}$, to the fermion-loop
result, while the bosonic isospin-breaking correction, $\De y^{\IB}_{\bos}$,
which is about an order of magnitude smaller than $\De y^{\SC}_{\bos}$
(see \refta{ta:delY}), is below experimental resolution. 

Combining our present result with previous ones~\cite{zph1,zph2} on 
$\De x$ and $\eps$, we find that the approximation
\beq
\De x\approx\De x_\fer, \qquad
\De y\approx\De y_\fer+\De y_\bos^\SC, \qquad
\eps\approx\eps_\fer, \quad
\label{eq:xyeapp}
\eeq
leads to results 
that deviate from the complete one-loop prediction for the effective
parameters by less than the experimental errors. 
In other words, the experimentally significant bosonic corrections are
completely contained in the effective parameter $\De y_\bos^\SC$ that
quantifies the effect of using the low-energy input parameter $\GF$ for
analyzing observables at the LEP energy scale. %From the discussion above
We recall that $\De y_\bos^\SC$, while being dependent on the
non-Abelian couplings of the gauge-bosons, is totally insensitive to
variations in the Higgs-boson mass.

In order to illustrate that the experimentally relevant bosonic
corrections to the LEP observables and the W-boson mass are indeed just
given by the contribution of $\De y_\bos^\SC$, we have compared 
in \reffis{sm3d}--\ref{mgfig} the experimental values of the observables
$\bar\sw^2$, $\Gl$, $\MWpm/\MZ$ with the corresponding
theoretical predictions. In \reffis{sm3d}a--\ref{mgfig}a the pure
fermion-loop predictions for various values of the top-quark mass,
$\Mt$, as well as the full one-loop results are compared with the
experimental data. 
In both theoretical predictions the leading two-loop
contributions of order $\ord(\alps\al t)$ and $\ord(\al^2t^2)$
have also been included (see \citere{zph2}).
Once the bosonic scale-change contribution, 
$\De y_\bos^\SC$, is added to the fermion-loop results, as shown in 
\reffis{sm3d}b--\ref{mgfig}b, there is complete agreement between theory
and experiment for values of the top-quark mass that are consistent with
the empirical value, $\Mt = 180 \pm 12 \GeV$~\cite{mtexp},
obtained from the direct search.
All other bosonic effects, in particular the %$\log(\MH)$-dependent
vacuum-polarization contributions contained in $\De x_\bos$ and
$\eps_\bos$, 
which show a logarithmic dependence for large values of the Higgs-boson
mass,
are below experimental resolution for Higgs-boson masses in
the perturbative regime, i.e.\ below $\sim 1\TeV$.

The experimental data used in \reffis{sm3d}--\ref{mgfig} 
read~\cite{data8/95}
\beqar
\Gl &=& 83.93 \pm 0.14 \MeV,\nn\\
\swbar^2 (\LEP) &=& 0.23186 \pm 0.00034, \nn\\
\frac{\MWpm}{\MZ} ({\mathrm UA2+CDF}) &=& 0.8802\pm0.0018. %\nn\\
\label{scdata}
\eeqar
We restrict our analysis to the LEP value of $\swbar^2$. Using
instead the combined LEP+SLD value, $\swbar^2=0.23143\pm0.00028$
\cite{data8/95}, does not significantly affect our results.
The theoretical predictions are based on
\beq
\MZ = 91.1884 \pm 0.0022 \GeV,
\label{scdataMZ}
\eeq
as well as the Fermi constant
\beq
\GF=1.16639 (2) \cdot 10^{-5} \GeV^{-2},
\label{eq:GF}
\eeq
the electromagnetic coupling at the $\PZ$-boson resonance,
\beq
\alpz^{-1} = 128.89 \pm 0.09,
\label{alpz1}
\eeq
which was taken from the recent updates~\cite{bu95} of the evaluation of
the hadronic vacuum polarization,
and the strong coupling
\beq
\alps(\MZ^2) = 0.123\pm0.006,
\label{scdataalps}
\eeq
also taken from \citere{data8/95}.

\newcounter{subfigure}
\setcounter{subfigure}{1}
\makeatletter
\def\thesubfigure{\@alph\c@subfigure}
\def\fnum@figure{\figurename~\thefigure\thesubfigure}
\newcommand{\sfcount}{\addtocounter{subfigure}{1}\addtocounter{figure}{-1}}
\makeatother
\begin{figure}
\begin{center}
\begin{picture}(16,13)
\put(0.7,6.5){$\bar\sw^2$}
\put(2,0.8){$\MWpm/\MZ$}
\put(11.4,0.6){$\Gl/\MeV$}
\put(-1.4,-6.0){\includegraphics{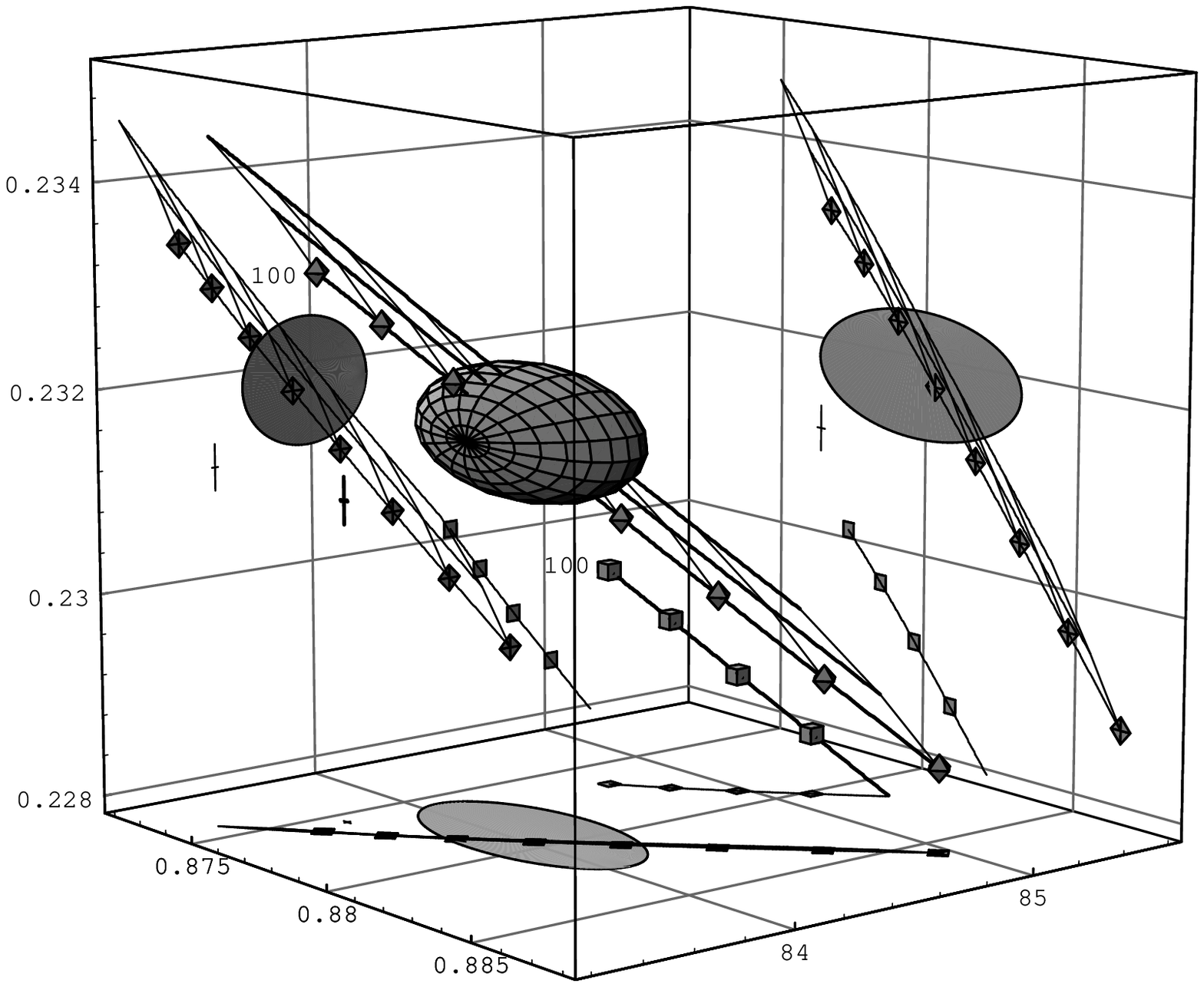}}
\end{picture}
\end{center}
\caption[]{Three-dimensional plot of the 68\% C.L.~($1.9 \sigma$) ellipsoid
of the experimental data in ($\MWpm/\MZ$, $\bar\sw^2$, $\Gl$)-space
and comparison with the full SM prediction (connected lines)
and the pure fermion-loop prediction (single line with cubes).
The full SM prediction is shown for Higgs-boson masses of $\MH = 100
\GeV$ (line with diamonds), $300\GeV$, and 
$1\TeV$ parametrized by $\Mt$
ranging from 100--240$\GeV$ in steps of $20\GeV$. In the pure
fermion-loop prediction the cubes also indicate steps in $\Mt$ of
$20\GeV$ starting with $\Mt = 100$ $\GeV$.
The cross outside the ellipsoid indicates the $\al(\MZ^2)$-Born
approximation with the corresponding error bars,
which also apply to all other theoretical predictions.
The projections onto the coordinate planes, which are included in the
plot, are also shown in \reffis{smfig}a--\ref{mgfig}a.}
\sfcount
\label{sm3d}
\efi
\begin{figure}
\begin{center}
\begin{picture}(16,13)
\put(0.7,6.5){$\bar\sw^2$}
\put(2,0.8){$\MWpm/\MZ$}
\put(11.4,0.6){$\Gl/\MeV$}
\put(-1.4,-6.0){\includegraphics{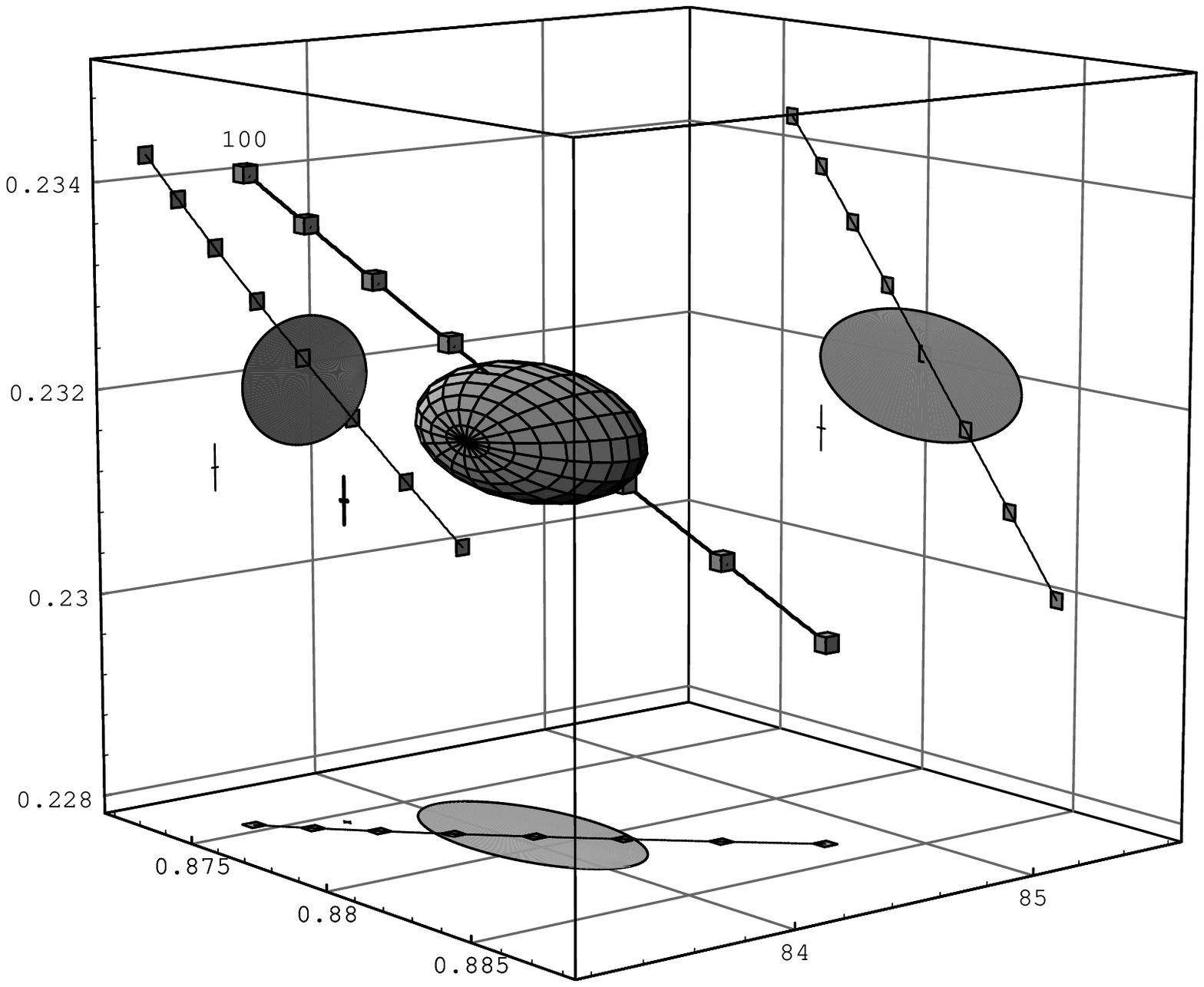}}
\end{picture}
\end{center}
\caption[]{Three-dimensional plot of the 68\% C.L.~($1.9 \sigma$) ellipsoid
of the experimental data in ($\MWpm/\MZ$, $\bar\sw^2$, $\Gl$)-space 
and comparison with the theoretical prediction obtained by combining
the fermion-loop contribution with the ($\MH$-independent)
bosonic correction $\De y_\bos^\SC$ related to the scale change from
$\GF$ to $\Ga^{\PW}_\Pl$ (compare \refeq{eq:delysc}). 
The theoretical prediction
is parametrized by $\Mt$ ranging from 100--240$\GeV$ in
steps of $20\GeV$.
The projections onto the coordinate planes, which are included in the
plot, are also shown in \reffis{smfig}b--\ref{mgfig}b.}
\efi
\makeatletter
\def\fnum@figure{\figurename~\thefigure}
\makeatother
\begin{figure}
\begin{center}
\begin{picture}(16,8.0)
\put(-2.7,-15.2){\includegraphics{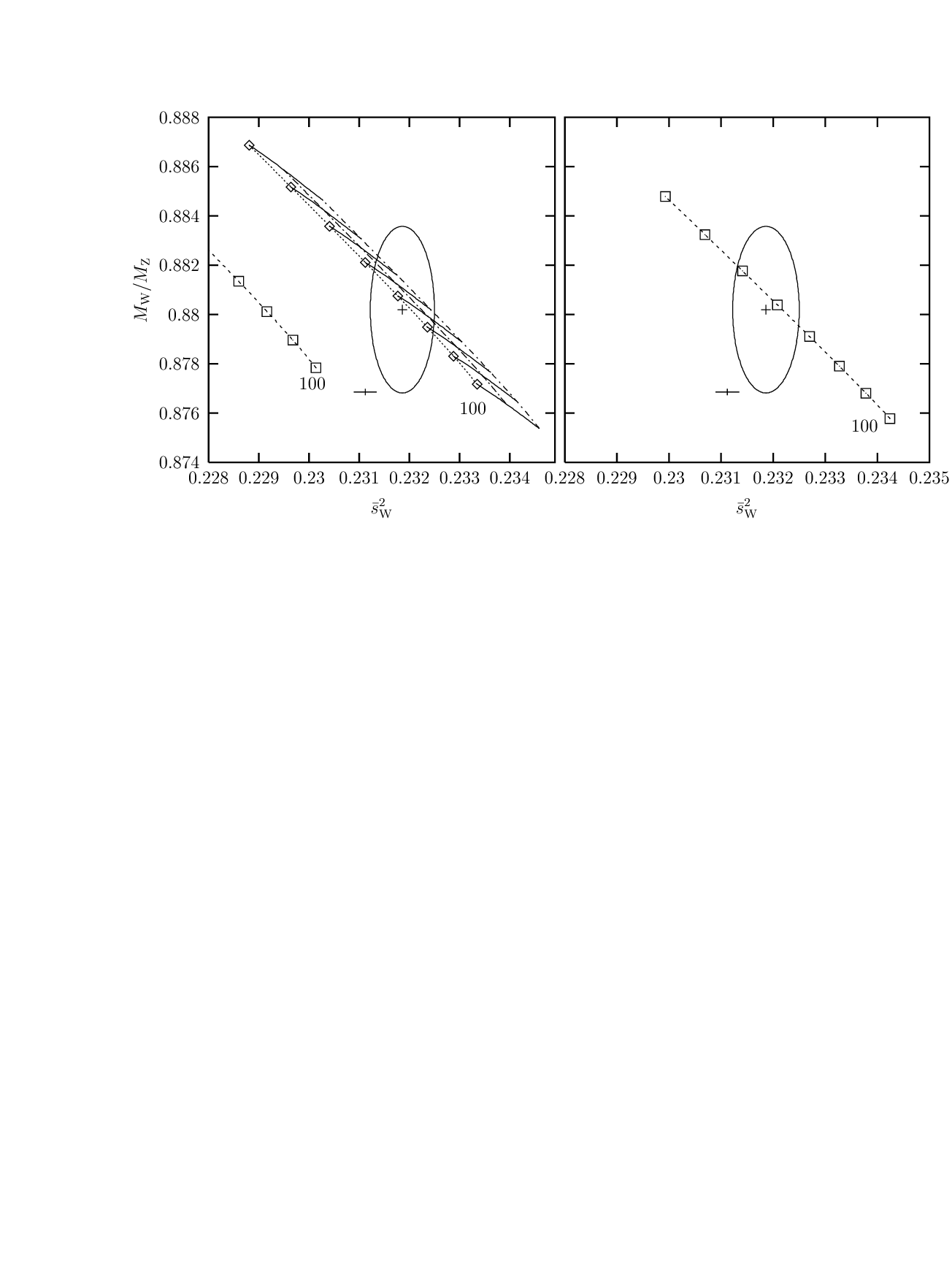}}
\put( 1.8,1.5){\footnotesize Fig.~\ref{smfig}a)}
\put( 8.8,1.5){\footnotesize Fig.~\ref{smfig}b)}
\end{picture}
\end{center}
\caption[]{Projection of the 68\% C.L.~($1.9 \sigma$) volume 
of the experimental data
in ($\MWpm/\MZ$, $\bar\sw^2$, $\Gl$)-space onto the
($\bar\sw^2$, $\MWpm/\MZ$)-plane.
In Fig.~\ref{smfig}a the
pure fermion-loop prediction is indicated by the
single line, the squares denote steps of $20\GeV$ in $\Mt$. The full SM
prediction is shown for Higgs-boson masses of $\MH = 100 \GeV$ (dotted
with diamonds), $300\GeV$ (long-dashed-dotted) and $1\TeV$
(short-dashed-dotted) parametrized by $\Mt$ ranging from
100--240$\GeV$ in steps of $20\GeV$. Fig.~\ref{smfig}b shows the 
theoretical prediction obtained by combining the 
fermion-loop contribution with the ($\MH$-independent)
bosonic correction $\De y_\bos^\SC$. 
}
\label{smfig}
\efi
\begin{figure}
\begin{center}
\begin{picture}(16,8.0)
\put(-2.7,-15.2){\includegraphics{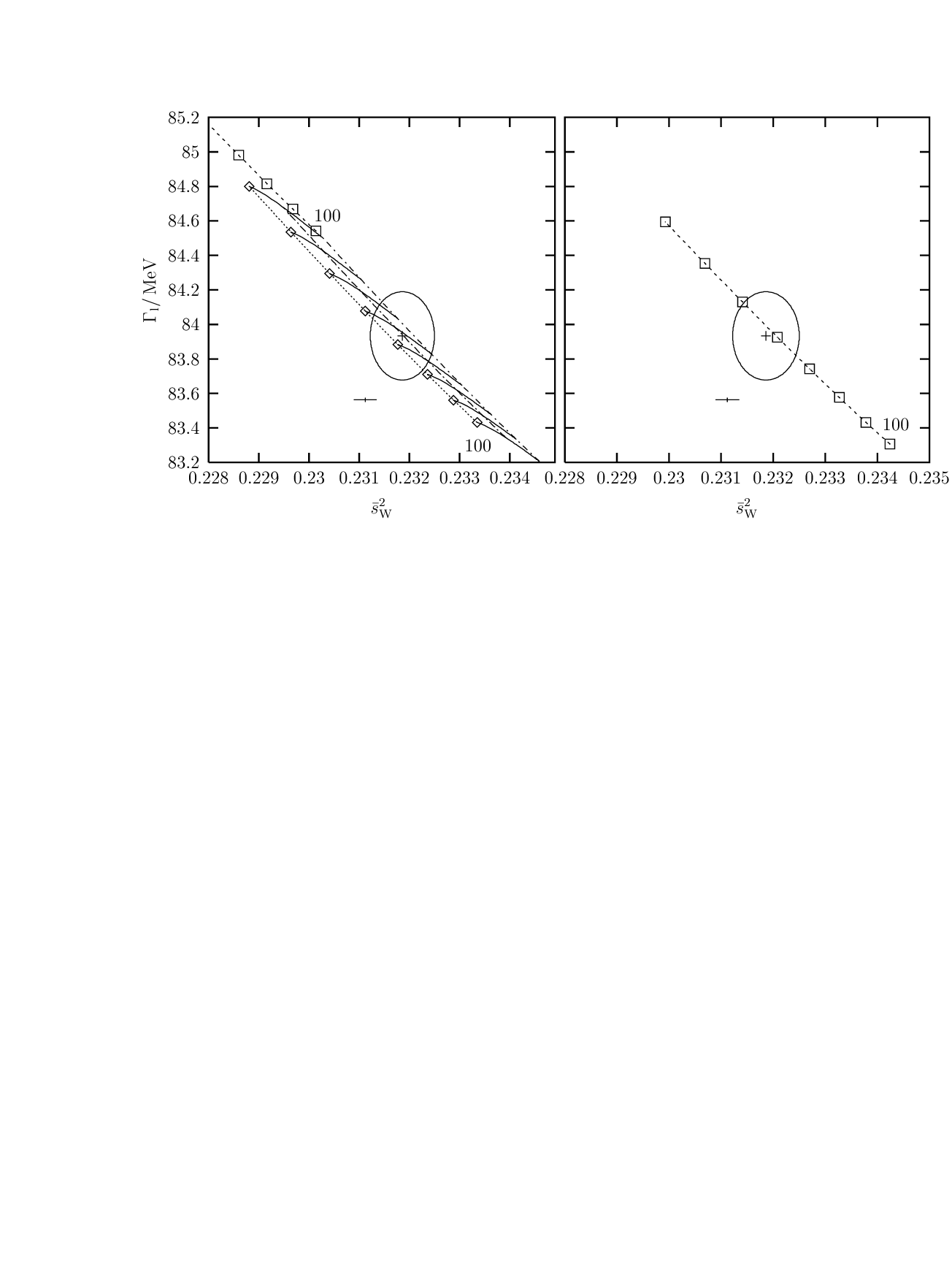}}
\put( 1.8,1.5){\footnotesize Fig.~\ref{sgfig}a)}
\put( 8.8,1.5){\footnotesize Fig.~\ref{sgfig}b)}
\end{picture}
\end{center}
\caption{Same signature as Fig.~\ref{smfig}, but for
the ($\bar\sw^2$, $\Gl$)-plane.}
\label{sgfig}
\efi
\begin{figure}
\begin{center}
\begin{picture}(16,8.0)
\put(-2.7,-15.2){\includegraphics{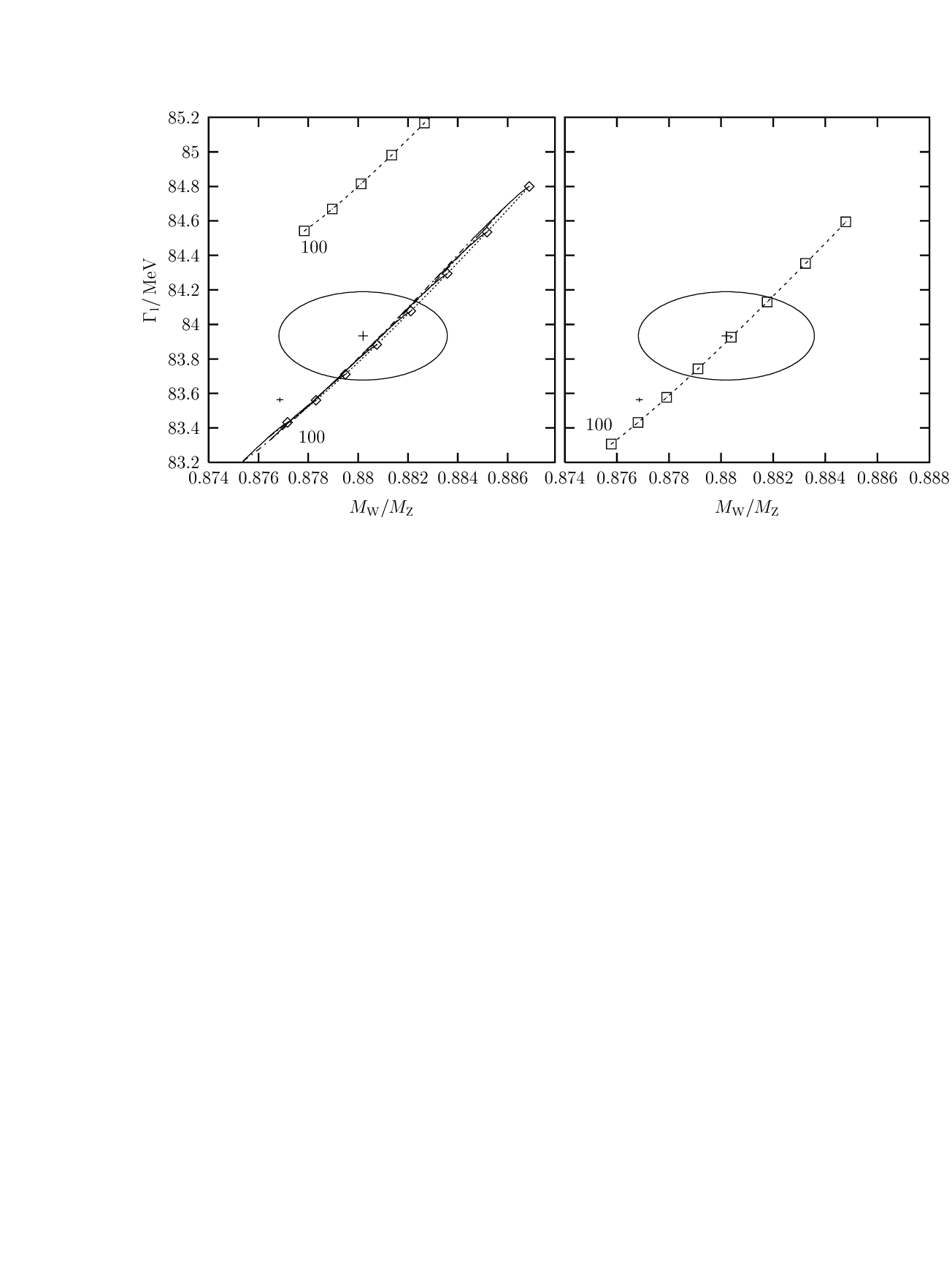}}
\put( 6.7,1.5){\footnotesize Fig.~\ref{mgfig}a)}
\put(14.0,1.5){\footnotesize Fig.~\ref{mgfig}b)}
\end{picture}
\end{center}
\caption{Same signature as Fig.~\ref{smfig}, but for
the ($\MWpm/\MZ$, $\Gl$)-plane.}
\label{mgfig}
\efi

In \reffis{sm3d}a and \ref{sm3d}b the 
volume defined by the data in the three-dimensional ($\MWpm/\MZ$,
$\bar\sw^2$, $\Gl$)-space corresponds to the 68\% C.L.~(i.e.~$1.9 \sigma$).
The projections of the 68\% C.L.~volume onto the planes of the
($\MWpm/\MZ$, $\bar\sw^2$, $\Gl$)-space correspond to
the 83\% C.L.\ ellipse in each plane, 
while the projections onto the individual axes correspond to the 94\%
C.L.\ there.
In addition to the three-dimensional plots in \reffi{sm3d},
the projections in each coordinate plane are also shown
in separate plots in \reffis{smfig}--\ref{mgfig}
for better illustration. Comparison of \reffis{smfig}a--\ref{mgfig}a 
with the corresponding plots of \citere{zph1}, where an analysis based
on the 1994 data was carried out, shows that the experimental error has
significantly decreased.

The fermion-loop predictions in \reffis{sm3d}a--\ref{mgfig}a 
are based on the
approximation
\beq
\De x\approx\De x_\fer, \qquad
\De y\approx\De y_\fer, \qquad
\eps\approx\eps_\fer. \quad
\eeq 
For $\Mt = 180 \GeV$ the pure fermion-loop predictions
at one loop read
\beqar
\bar s^2_{{\ss\PW}, {\fer}} &=& 0.22747 \mp 0.00023, \nn\\
\left(\frac{\MWpm}{\MZ}\right)_{\fer} &=& 0.88358 \pm 0.00013, \nn\\
\Ga_{\Pl, {\fer}} &=& 85.299 \pm 0.012\MeV, 
\label{eq:ferpred}
\eeqar
which deviate from the experimental values by $-13\si$, 
$1.9\si$ and $9.8\sigma$, respectively.

The uncertainties of the theoretical predictions
are dominated by the error of $\alpz$ given in \refeq{alpz1}.
In \reffis{sm3d}--\ref{mgfig} the $\alpz$-Born approximation,
\beq
s_0^2 = 0.23112 \mp 0.00023, \quad c_0 = 0.87686 \pm 0.00013, \quad
\Gl^{(0)} = 83.563 \pm 0.012\MeV, \quad
\label{eq:alpzborn}
\eeq
is also indicated for completeness. The error bars shown for the 
$\alpz$-Born approximation  of course also apply to the other theoretical
predictions.
The $\alpz$-Born approximation corresponds to a deviation of $-2.2\si$, 
$-1.9\si$ and $-2.6\si$, respectively, from the experimental data.
The fact that the values in \refeq{eq:alpzborn} are closer to the empirical
data and the full SM predictions than the fermion-loop prediction
\refeq{eq:ferpred} is a consequence of the cancellation
between fermionic and bosonic contributions in the single parameter 
$\De y^\SC$ as displayed in \refta{ta:delY}.

The comparison of the full SM predictions with the data in 
\reffis{sm3d}a--\ref{mgfig}a also 
illustrates in how far the data are sensitive to variations in the 
Higgs-boson mass.
For fixed values of $\Mt \sim 180 \GeV$ 
the intersection of the 68\% C.L.~($1.9 \sigma$) volume with 
the SM prediction allows for Higgs masses which span the full range from
the experimental lower bound of $\sim 60 \GeV$ to about $1 \TeV$.
It should also be noted that in the direction in three-dimensional space
in which the $\MH$-dependence (for fixed $\Mt$) is sizable also the
uncertainty in the theoretical predictions due to the error in
$\al(\MZ^2)$ is large.

\section{\boldmath{Radiative corrections in the 
$\Ga^{\PW}_{\mathrm l}$-scheme}}
\label{sec:gawscheme}

In the previous section,
using the framework of an effective Lagrangian and the specific
parametrization of the radiative corrections in terms of $\De x$, $\De
y$, and $\eps$, we have identified the experimentally relevant bosonic
corrections in the analysis of the precision data with the contribution
of the single effective parameter $\De y_\bos^\SC$. In the present
section we will formulate this observation in a way that is 
independent of any specific parametrization and that makes the physical
interpretation of those bosonic corrections that are in fact tested in 
current precision experiments more transparent.

As shown in the last section, the parameter $\De y_\bos^\SC$ expresses
the effect of using the low-energy quantity $\GF$ as input parameter for
the analysis of the LEP observables. We will now demonstrate that these
large bosonic corrections could indeed completely be avoided by
expressing 
the theoretical predictions for the observables $\bar\sw^2$,
$\Gl$, $\MWpm/\MZ$ in terms of appropriate 
input parameters being defined at
the scale of the vector-boson masses, namely
by using the $\PW$-boson
width $\Ga^{\PW}_{\mathrm l}$ instead of the Fermi constant $\GF$
as input parameter.

In the language of the effective Lagrangian ${\cal L}_{\mathrm C}$ given
in \refeq{lc} the use of the input quantity $\Ga^{\PW}_{\mathrm l}$ instead
of $\GF$ means that the charged-current coupling in \refeq{lc}
is identified with $g_\PWpm(\MWpm^2)$ defined via the $\PW$-boson width
$\Ga^{\PW}_{\mathrm l}$ (see \refeq{eq:Wlepwidth}) rather than with 
$g_\PWpm(0)$ defined via muon decay (see \refeq{eq:gw0}).
With this identification
the transition between $g_\PWpm(0)$ and $g_\PWpm(\MWpm^2)$, and
accordingly the contribution of $\De y^\SC$, does not occur. The
radiative corrections to the observables 
$\swbar^2$, $\Gl$, and $\MWpm/\MZ$ are completely contained in the
parameters $\De x$, $\De y^{\IB}$, and $\eps$, in which SM
corrections beyond fermion loops do not give rise to significant
contributions.

In this ``$\Ga^\PW_\Pl$-scheme'' the lowest-order values
$\hat s_0^2$, $\hat c_0$, and $\hat \Gl^{(0)}$ of the 
observables are given in terms
of the input quantities $\al (\MZ^2)$, $\MZ$, and $\Ga^\PW_\Pl$
as
\beq
\frac{\hat s_0^2}{\hat c_0} \equiv \frac{\al (\MZ^2) \MZ}{12
\Ga^{\PW}_{\mathrm l}} \left(1 + c_0^2 \frac{3 \al}{4 \pi}\right),
\qquad \hat c_0^2 \equiv (1 - \hat s_0^2),
\eeq
and 
\beq
\hat\Gl^{(0)} = \frac{\alpz\MZ}{48 \hat s_0^2 \hat c_0^2}
\left[1+(1-4 \hat
s_0^2)^2\right]\left(1+\frac{3\alpha}{4\pi}\right) .
\eeq
The relations between the observables and the effective
parameters $x$, $y^\IB$, and $\eps$ read
\beqar
\swbar^2 \left( 1-\swbar^2 \right)
& = &
\frac{\hat s_0^2}{\hat c_0} \frac{\MWpm^3} {\MZ^3}
\frac {y^{\IB} - 2 \hat s_0^2 \de}{x + 2 \hat s_0^2 \de}
(1-\eps + \de)
\frac{1} {\left( 1 +\frac{\bar\sw^2} {1-\bar\sw^2}
(\eps  - \de)\right)}, \nn\\
\frac {\MWpm^2} {\MZ^2}
& = &
\left( 1-\bar\sw^2 \right) \left(x + 2 \hat s_0^2 \de \right)
\left( 1 + \frac{\bar\sw^2}{1-\bar\sw^2}(\eps - \de)\right),
\nn\\
\Gl
& = &
\frac {\Ga^{\PW}_{\mathrm l} \MZ^3} {4 \MWpm^3}
\left[ 1 + \left( 1 - 4\bar\sw^2\right)^2\right]
\frac{x + 2 \hat s_0^2 \de}{y^{\IB} - 2 \hat s_0^2 \de}
\left( 1 + s_0^2 \frac{3\alpha}{4\pi}\right).
\label{eq:analW}
\eeqar
The small parameter $\de$ ($\de \sim 10^{-4}$ in the SM),
which describes parity violation in the photonic coupling at the
Z-boson mass scale, has been defined in \citere{zph2}.
The relations \refeq{eq:analW} are simply obtained from eqs.~(16)
of \citere{zph2}, where the observables are expressed in terms
of the input parameters $\al (\MZ^2)$, $\MZ$, and $\GF$, by the
replacement %of $\GF/y^\SC$ via
\beq
\GF = y^\SC \Ga^{\PW}_{\Pl}  \frac{6 \sqrt{2} \pi}{\MWpm^3}
\left(1 + c_0^2 \frac{3 \al}{4 \pi}
\right)^{-1} ,
\eeq
which explicitly illustrates that the contribution of $\De y^\SC$ is
absorbed by introducing the ``large-scale'' quantity
$\Ga^{\PW}_{\mathrm l}$.
Linearizing~\refeq{eq:analW} in $\De x$, $\De y^{\IB}$,
$\eps$, and $\de$ yields 
\beqar
\bar\sw^2 &=& \hat s_0^2\left[1+
\frac{\hat c_0^2}{2- \hat s_0^2} \De x +
\frac{2 \hat c_0^2}{2- \hat s_0^2} \De y^{\IB} +
\frac{3\hat s_0^2 - 2}{2- \hat s_0^2} \eps
+ (\hat c_0^2 - \hat s_0^2) \de \right], \nn\\
\frac{\MWpm}{\MZ} &=& \hat c_0\left[1+
\frac{\hat c_0^2}{2- \hat s_0^2} \De x -
\frac{\hat s_0^2}{2- \hat s_0^2} \De y^{\IB} +
\frac{2 \hat s_0^2}{2- \hat s_0^2} \eps
\right], \nn\\
\Gl &=& \hat \Gl^{(0)}\biggl[1 -
\frac{2}{(2- \hat s_0^2) \left[1 + (1 - 4 \hat s_0^2)^2 \right]}
\left((1 - 2 \hat s_0^2 - 4 \hat s_0^4) (\De x + 2 \De y^{\mathrm
IB})
\right. \nn\\
&& \left. {} \qquad \qquad
- 2 \hat s_0^2 (1 - 10 \hat s_0^2) \eps - 8 \hat s_0^4 (2- \hat
s_0^2) \de
\right) \biggr] .
\label{eq:analWlin}
\eeqar

The relations \refeq{eq:analWlin} could in principle be used for
a data analysis of the observables $\bar\sw^2$, $\Gl$, and
$\MWpm/\MZ$ in the $\Ga^{\PW}_{\Pl}$-scheme, i.e.~with
$\al(\MZ^2)$, $\MZ$, and $\Ga^{\PW}_{\Pl}$ as experimental input
quantities. Assuming (hypothetically) the same experimental
accuracy as in the ``$\GF$-scheme'' (input parameters
$\al(\MZ^2)$, $\MZ$, and $\GF$) and an experimental value of
$\Ga^{\PW}_{\Pl}$ being in agreement with the SM prediction,
a consistent description of the
data in the $\Ga^{\PW}_{\Pl}$-scheme would be possible by solely
including the pure fermion-loop predictions in the effective
parameters.

At present a data analysis using the $\Ga^{\PW}_{\Pl}$-scheme
would of course not be sensible owing to the large experimental
error in the determination of the $\PW$-boson width. {}From
\citere{PDB} we have $\Ga^{\PW,\mathrm exp}_{\mathrm T} = (2.08 \pm
0.07) \GeV$ for the total decay width of the W-boson and $(10.7
\pm 0.5) \%$ for the leptonic branching ratio. Adding the errors
quadratically yields $\Ga^{\PW,\mathrm exp}_{\mathrm l} = (223 \pm 13)
\MeV$ showing that the experimental error in $\Ga^{\PW,\mathrm
exp}_{\mathrm l}$ at present is more than an order of magnitude larger
than the error in the leptonic \PZ-boson width (see
\refeq{scdata}) and obviously much larger than the one in $\GF$
(see \refeq{eq:GF}).

Even though a precise experimental input value for $\Ga^\PW_\Pl$
is not available it is nevertheless instructive to simulate the
analysis in the $\Ga^\PW_\Pl$-scheme by using the theoretical SM
value for $\Ga^\PW_\Pl$ as hypothetical input parameter for
evaluating \refeq{eq:analWlin}. For the choice of $\Mt = 180
\GeV$ and $\MH = 300 \GeV$ one obtains 
$\Ga^{\PW}_{\Pl} = 226.3 \MeV$
as theoretical value of $\Ga^\PW_\Pl$ in the SM.
One should note that our procedure here is technically analogous
to commonly used parametrizations of radiative corrections
where, for instance in
the on-shell scheme (see e.g.~\citere{adhabil}), the 
corrections are expressed in terms of the $\PW$-boson mass 
$\MWpm$, while in an actual evaluation $\MWpm$ is substituted by
its theoretical SM value in terms of $\alpz$, $\MZ$, and $G_\mu$. 

\btab 
\renewcommand{\arraystretch}{1.7}
\arraycolsep 5pt
$$\begin{array}{|c||c|c||c|c||c|}
\hline
& \multicolumn{2}{|c||}{\GF{\mathrm -scheme}} &
  \multicolumn{2}{c||}{\Ga^{\PW}_{\mathrm l}{\mathrm -scheme}} & 
\\ \cline{2-5}
& {\mathrm ferm.~corr.} & {\mathrm bos.~corr.} & 
  {\mathrm ferm.~corr.} & {\mathrm bos.~corr.} & 
  {\mathrm exp.~error} \\ \hline\hline
\frac{\De\bar\sw^2}{\bar\sw^2}/ 10^{-3} & 
-15.8 & \phantom{-}16.3 & \phantom{-}11.0 & \phantom{-}1.3 & 
1.5 \\ \hline
\frac{\De\MWpm}{\MWpm} /10^{-3} & 
\phantom{-1}7.7 & ~-1.6 & \phantom{-1}3.7 & 
\phantom{-}0.6 & 2.0 \\ \hline
\frac{\De \Gl}{\Gl} / 10^{-3} & 
\phantom{-}20.8 & -14.3 & ~-1.8 & -1.7 & 1.7 \\ \hline
\earr$$
\caption{
Relative size of the SM fermionic and bosonic one-loop corrections
to the observables $\bar\sw^2$, $\MWpm$, and $\Gl$ in the $\GF$-scheme
(input parameters $\GF$, $\MZ$, and $\al (\MZ^2)$)
and in the simulated $\Ga^{\PW}_{\mathrm l}$-scheme 
(input parameters $\Ga^{\PW}_{\Pl} = 226.3 \MeV$, $\MZ$, 
and $\al (\MZ^2)$)
for  $\Mt = 180 \GeV$ and $\MH = 300 \GeV$.
The relative experimental error of the observables is also indicated.}
\label{tab:analW}
\etab

In order to illustrate the fact that the replacement of the input
quantity $\GF$ by $\Ga^{\PW}_{\mathrm l}$ indeed strongly affects the
relative size of the fermionic and bosonic contributions
entering each observable,
we have given in \refta{tab:analW} the 
relative values of the SM one-loop fermionic and bosonic
corrections to the observables $\bar\sw^2$, $\MWpm$, and $\Gl$
in the $\GF$-scheme and in the simulated $\Ga^{\PW}_{\mathrm
l}$-scheme based on the input value $\Ga^{\PW}_{\Pl} = 226.3
\MeV$. The size of the radiative corrections in the two schemes
is compared with the relative experimental error of the observables
(see \refeq{scdata}).
Table \ref{tab:analW} shows that in the $\GF$-scheme the bosonic
corrections to $\bar\sw^2$ and $\Gl$ are quite sizable and considerably
larger than the experimental error. In the (simulated)
$\Ga^{\PW}_{\mathrm l}$-scheme,
on the other hand, these corrections are smaller by 
an order of magnitude
and have about the same size as the experimental
error. The bosonic contributions to $\MWpm$ are smaller than the
experimental error in both schemes. 
It can furthermore be seen in \refta{tab:analW} that the
cancellation between fermionic and bosonic corrections related
to the scale change is not present in the $\Ga^{\PW}_{\Pl}$-scheme.

The explicit values for the pure fermion-loop predictions of the
observables 
at one-loop order
in the simulated $\Ga^{\PW}_{\Pl}$-scheme read
\beq
\bar s^2_{{\ss \PW}, {\fer}} = 0.23154, \qquad
\left(\frac{\MWpm}{\MZ}\right)_{\fer} = 0.8813, \qquad
\Ga_{\Pl, {\fer}} = 84.06 \MeV. \quad
\eeq
Comparison with the experimental values of the observables given
in \refeq{scdata} shows that there is indeed no significant
deviation between the pure fermion-loop predictions in the
simulated $\Ga^{\PW}_{\Pl}$-scheme and the data,
i.e.\ they agree within one standard deviation. This has to be
contrasted to the situation in the $\GF$-scheme, where the pure
fermion-loop predictions differ from the data by several
standard deviations (see \refeq{eq:ferpred} and
\reffi{sm3d}a--\ref{mgfig}a).

In summary, we have demonstrated that after replacing 
the low-energy quantity $\GF$ by the high-energy observable
$\Ga^{\PW}_{\Pl}$ in the theoretical predictions 
for the observables $\bar\sw^2$, $\MWpm/\MZ$, and $\Gl$,
no corrections beyond fermion loops are required in order to
consistently describe the data. Although at present,
due to the large experimental error in $\Ga^{\PW}_{\Pl}$, 
the so-defined $\Ga^{\PW}_{\Pl}$-scheme is 
of no practical use for analyzing these precision data, 
from a theoretical point of view
it shows that the only bosonic corrections of
significant magnitude 
are such that they can completely be absorbed by
the introduction of the quantity $\Ga^{\PW}_{\Pl}$.
It is obvious that all results of this section do not depend on any
specific parametrization, such as the description via $\De x$, 
$\De y^\IB$, $\eps$, chosen for analyzing the data. 
They only rely on the set of physical observables chosen as input
parameters.

\section{Conclusions}
\label{conc}

It is by now a well established fact that the experimental data on 
the leptonic LEP1 observables $\Gl$, $\bar\sw^2$ and the $\PW$-boson 
mass $\MWpm$ are sensitive to radiative corrections beyond pure
fermion loops. In this paper we have investigated in detail
which standard electroweak bosonic loop corrections %that 
are in fact tested by the experiments.

Starting from an analysis of the radiative corrections in terms of the
effective parameters $\De x$, $\De y$, and $\eps$, which quantify
different sources of SU(2) violation in an effective Lagrangian, we have
shown that the bosonic corrections needed for an agreement between the
SM predictions and the current precision data can be identified as an
effect of the transition from the low-energy process muon decay to the
LEP observables. In the framework of the effective Lagrangian all
experimentally significant bosonic corrections are contained in the
single effective parameter $\De y^\SC$ that quantifies the difference
between the charged-current coupling $g_\PWpm(0)$ 
defined via the low-energy process
muon decay
and the charged-current coupling $g_\PWpm(\MWpm^2)$ at the W-boson mass
shell. This fact has been demonstrated not only at the level of the effective
parameters, but also by confronting theory and experiment
for the observables $\Gl$, $\bar\sw^2$, and $\MWpm$ themselves.

As a consequence of these investigations, a simple physical
interpretation of those bosonic corrections that are resolved
by the current precision experiments can be given. They
are precisely the ones that appear when W-boson decay
is expressed in terms of the input parameter $\GF$ 
and can therefore directly 
be related to the set of physical observables chosen as input 
parameters for the data analysis.
Upon introducing the high-energy quantity $\Ga^{\PW}_{\Pl}$ 
($\Ga^{\PW}_{\Pl}$-scheme) as input parameter instead of
$\GF$ ($\GF$-scheme), the significant bosonic corrections are
completely absorbed when analyzing the 
high-energy observables $\swbar^2$, $\Gl$, and $\MWpm/\MZ$.
Since the usefulness of $\Ga^{\PW}_{\Pl}$ as experimental input parameter
is limited at present due to the large experimental error of the
W-boson width, we have demonstrated this fact by invoking the SM
theoretical value of $\Ga^{\PW}_{\mathrm l}$ as input. Indeed, no further
corrections beyond fermion loops are needed in this case in order to
achieve agreement with the data within one standard deviation.

It is amusing to note that the precision data collected at LEP, i.e.\
for a neutral-current process at the Z-boson resonance,
provide a test of just those bosonic radiative corrections that are
associated with the charged-current process of W-boson decay.
While the experimentally significant bosonic corrections are insensitive to
variations in the Higgs-boson mass, they provide a highly non-trivial, even
though indirect, test of the non-Abelian gauge structure of the
electroweak theory.

\section*{Acknowledgement}
The authors thank Masaaki Kuroda for interesting discussions.
One of the authors (D.S.) thanks the Max Planck Institut, Werner
Heisenberg Institut f\"ur Physik, M\"unchen, for hospitality while part
of the present work was done.

\appendix
\def\theequation{\thesection.\arabic{equation}}
\setcounter{equation}{0}
\section*{Appendix}
\section{\boldmath{Remainders of $\De y^\SC_\bos$ and $\De y^\IB_\bos$}}
\label{sec:remain}

In \refse{sec:dely} we have split the parameters $\De y^{\SC,\IB}_\bos$ 
into dominant parts $\De y^{\SC,\IB}_\bos({\mathrm dom})$ and 
remainder parts containing the $\MH$-dependence, which is numerically
completely negligible. The remainder parts read explicitly
\beqar
\label{eq:delyscbosrem}
\Delta y^\SC_\bos({\mathrm rem}) &=& 
\frac{\al(\MZ^2)}{32 \pi c_0^6 s_0^2}
\Biggl[ - \frac{c_0^2}{3} (47 c_0^4 - 21 c_0^2 h + 6 h^2) \nn\\
&& {} + \frac{1}{c_0^2 - h}
(4 c_0^8 - 22 c_0^6 h + 17 c_0^4 h^2 - 6 c_0^2 h^3 + h^4) 
\log\left(\frac{c_0^2}{h}\right) \nn\\
&& {} + \frac{1}{4 c_0^2 - h} h (28 c_0^6 - 20 c_0^4 h + 7 c_0^2 h^2 - h^3)
f_2\left(\frac{c_0^2}{h}\right) \Biggr],
\\[.5em]
\label{eq:delyibbosrem}
\Delta y^\IB_\bos({\mathrm rem}) &=& 
\frac{\al(\MZ^2)}{96\pi s_0^2} \Biggl[ 
-\frac{4s_0^2h}{c_0^4}(3c_0^2-h-c_0^2h)
\nn\\ && {}
+\left(10-18\frac{h}{c_0^2}+9\frac{h^2}{c_0^4}-2\frac{h^3}{c_0^6}\right)
\log\left(\frac{h}{c_0^2}\right)
-(10-18h+9h^2-2h^3)\log(h)
\nn\\ && {}
-\left(36-32\frac{h}{c_0^2}+13\frac{h^2}{c_0^4}-2\frac{h^3}{c_0^6}\right)
\frac{h}{4c_0^2-h}\logy\left(\frac{c_0^2}{h}\right)
\nn\\ && {}
+(36-32h+13h^2-2h^3)\frac{h}{4-h}\logy\left(\frac{1}{h}\right)
\Biggr],
\eeqar
where the shorthand $h = \MH^2/\MZ^2$ is used. By definition,
$\De y^{\SC,\IB}_\bos({\mathrm rem})$ approach asymptotically zero
for $h \to \infty$.

\section{Auxiliary functions}
\label{aux}

Here we list the explicit expressions for the auxiliary functions
which have been used in \refse{sec:dely} and \refapp{sec:remain}
for the explicit
formulae of the bosonic contributions
to $\De y^\SC$ and $\De y^\IB$. As defined
in \citere{zph1}, $f_{1,2}(x)$ are given by
\beqar
\logx(x) &=& \Re\left[\beta_x\log\left(\frac{\beta_x-1}
              {\beta_x+1}\right)\right], \quad\mbox{with}\;\;
             \beta_x=\sqrt{1-4x+i\epsilon}, \nn\\
\logy(x) &=& \Re\left[\beta_x^*\log\left(\frac{1-\beta_x^*}
              {1+\beta_x^*}\right)\right]. 
\eeqar
Furthermore, some scalar one-loop three-point integrals 
have been abbreviated by
\beqar
C_1 &=& \MZ^2\Re\left[C_0(0,0,\MZ^2,0,\MZ,0)\right] =
-\frac{\pi^2}{12} = -0.8225, \nn\\[.3em]
C_2 &=& \MZ^2\Re\left[C_0(0,0,\MZ^2,0,\MW,0)\right] =
\frac{\pi^2}{6}-\Re\left[\Li\left(1+\frac{1}{c_0^2}\right)\right]
= -0.8037, \nn\\[.3em]
C_3 &=& \MZ^2\Re\left[C_0(0,0,\MZ^2,\MW,0,\MW)\right]
= \Re\left[\log^2\left(\frac{i\sqrt{4c_0^2-1}-1}
                          {i\sqrt{4c_0^2-1}+1}\right)\right]
= -1.473, \nn\\
C_6 &=& c_0^2\MZ^2\Re\left[C_0(0,0,\MWpm^2,0,\MZ,0)\right] =
\frac{\pi^2}{6}-\Re\left[\Li\left(1+c_0^2\right)\right]
= -0.8067, \nn\\[.3em]
C_7 &=& c_0^2\MZ^2\Re\left[C_0(0,0,\MWpm^2,\MWpm,0,\MZ)\right] \nn\\
&=&
-\log\left[\frac{1}{2}\left(1+i\sqrt{4c_0^2-1}\right)\right]
 \log\left[\frac{1}{2}\left(1-i\sqrt{4c_0^2-1}\right)\right]
= -0.9466,
\eeqar
with the dilogarithm
\beq
\Li(x) = -\int_0^x\,\frac{dt}{t}\,\log(1-t),
\qquad -\pi<{\mathrm arc}(1-x)<\pi.\\[.2em]
\eeq
The first three arguments of the $C_0$-function label the external
momenta
squared, the last three the inner masses of the corresponding vertex
diagram.

\end{document}